\documentclass[preprint]{aastex}

\slugcomment{To appear in the Astrophysical Journal}
\received{5/24/2000}
\revised{8/21/2000}
\accepted{10/3/2000}

\newcommand{\iras}{{\sl IRAS\/ }}
\newcommand{\iso}{{\sl ISO\/ }}
\newcommand{\um}{\mu m}
\newcommand{\mdot}{{\dot M}}
\newcommand{\msun}{M_{\odot}}
\newcommand{\lsun}{L_{\odot}}
\newcommand{\rsun}{R_{\odot}}
\newcommand{\rstar}{R_{*}}
\newcommand{\lstar}{L_{*}}
\newcommand{\teff}{T_{\rm eff}}
\newcommand{\tdust}{T_{\rm d}}
\newcommand{\rdust}{r_{\rm d}}

\newcommand{\fstarobs}{F_{*}^{\rm obs}}
\newcommand{\fstar}{F_{*}}
\newcommand{\qabs}{Q_{\rm abs}}

\newcommand{\crich}{C$-$rich~}
\newcommand{\orich}{O$-$rich~}
\makeatletter
\def\ale{\mathrel{\mathpalette\gl@align<}}
\def\age{\mathrel{\mathpalette\gl@align>}}
\def\gl@align#1#2{\lower.6ex\vbox{\baselineskip\z@skip\lineskip\z@
\ialign{$\m@th#1\hfil##\hfil$\crcr#2\crcr\sim\crcr}}}
\makeatletter

\begin{document}
 
\title{Discovery of an Extended Dust Emission around 
IRAS 18576$+$0341
(AFGL 2298) at 10.3 and 18.0 microns:
a New Luminous Blue Variable Candidate?}

\author{Toshiya Ueta and Margaret Meixner}
\affil{Department of Astronomy, MC-221, 
University of Illinois at Urbana-Champaign, 
Urbana, IL  61801,
ueta@astro.uiuc.edu, meixner@astro.uiuc.edu}

\author{Aditya Dayal}
\affil{IPAC/JPL, Caltech, MS 100-22,
770 South Wilson Ave. 
Pasadena, CA  91125, 
adayal@ipac.caltech.edu}

\author{Lynne K. Deutsch}
\affil{Department of Astronomy/CAS 519, 
Boston University, 
725 Commonwealth Avenue, 
Boston, MA  02215,
deutschl@bu.edu}

\author{Giovanni Fazio and Joseph L. Hora}
\affil{Harvard-Smithsonian Center for Astrophysics, MS 65,
60 Garden St., 
Cambridge, MA  02138,
jhora@cfa.harvard.edu, gfazio@cfa.harvard.edu}

\author{William F. Hoffmann}
\affil{Steward Observatory, University of Arizona, 
Tucson, AZ  85721, 
whoffmann@as.arizona.edu}

\begin{abstract}
We report detection of an extended mid-infrared emission from \iras 
18576+0341 (AFGL 2298).
The object shows a dusty circumstellar shell that
has diameter of $\age 7\arcsec$ at 10.3 and 18.0 $\um$.
The dust nebula shows two emission peaks concentrically elongated
and symmetrically oriented on the opposite sides of the third,
central peak, which appears to be the central star of the
system.
The observed mid-infrared morphology indicates that the
circumstellar dust shell has an equatorially-enhanced material
distribution, which is a common signature of stellar objects
that have experienced mass loss.
Radiative transfer model calculations suggest that the
central star is an extremely bright ($\lstar = 10^{6.4} \lsun$)
star at a distance of about 10 kpc:
this object is best described as a new luminous blue variable
candidate.
The circumstellar dust shell seems to have been generated by 
an equatorially-enhanced mass loss process with 
${\dot M} \ge 6.8 \times 10^{-6} \msun$ yr$^{-1}$ and
${\dot M}_{\rm pole}/{\dot M}_{\rm eq} \sim 0.5$.
\end{abstract}


\section{Introduction}

Mid-infrared (mid-IR; $8 - 25 \um$) imaging provides powerful 
means to directly probe the distribution of circumstellar matter, 
especially circumstellar dust grains.
High-resolution ($\sim 1\arcsec$) mid-IR images can reveal the 
structure at the innermost regions of the circumstellar dust shells 
(CDSs), because the emission is generally optically thin and 
arises from the warmest ($100 - 200$ K) dust grains in the CDSs. 
Such mid-IR images of the CDSs have shown that many stars have 
lost or have been losing their mass in an equatorially-enhanced 
manner, resulting in toroidal or ring shaped CDSs.
For example, the axisymmetric shaping of planetary nebulae is
considered to be initiated by an equatorially-enhanced
dust-driven wind mass loss at the end of the asymptotic giant 
branch phase (e.g., \citealt{ueta00} and references therein).
The axisymmetric morphologies of the nebulae around luminous 
blue variables (LBVs) are explained similarly by the presence of
an equatorially-enhanced CDS \citep{nota95}, which may be 
generated by stellar rotation 
near the Eddington limit \citep{langer99} or the tidal action
of the central binary system \citep{damineli00}.

In this paper, we report our discovery of an extended 
dust shell around an unidentified IR object, 
\iras 18576+0341 (\iras 18576, hereafter),
during our observing run to obtain high-resolution
mid-IR images of CDSs around evolved stars.
After reviewing previous observations in \S 2,
we present the observed images in \S 3.
The results of the dust radiative transfer model calculations 
are presented in \S 4, followed by a discussion on the nature 
of the source in \S 5.
Finally, our conclusion is presented in \S 6.

\section{The Object: IRAS 18576+0341}

\iras 18576 was first detected as an IR source by 
the Survey Program of Infrared Celestial Experiments sensor and the 
Far-Infrared Sky Survey Experiment telescope \citep{pms81}.
Their successor program, the Air Force Geophysical Laboratory (AFGL) 
Infrared Sky Survey \citep{pm83}, confirmed the detection
of this IR source and designated the object as AFGL 2298.
Infrared Astronomical Satellite ({\it IRAS\/}; \citealt{iras}) 
also detected it as a point source and assigned the
\iras ID (\iras 18576+0341).
From its \iras color, \iras 18576 was classified as a
region V object \citep{vh88} and was suspected to be a
planetary nebula (PN), 
which is consistent with the source's very red nature 
(class H) revealed by the \iras Low Resolution Spectra 
(LRS; \citealt{vc89, kvb97}).
The LRS also suggested the oxygen-rich (O$-$rich) nature of this object
due to its possible silicate absorption feature at $10 \um$.
\citet{gmpp97} obtained near-IR photometric data of the source 
and classified it as a post-AGB object.
However, no optical counterpart has been identified
(\citet{hvk00} report their unpublished photometry of V $>$ 23.2).

\citet{zd86} were unsuccessful in their search for CO line
emission toward \iras 18576 with the National Radio Astronomical 
Observatory (NRAO) 12 m telescope.
The authors, nevertheless, classified the object as an 
oxygen-rich source based on its \iras color.
No detection of 22.2 GHz H$_2$O maser emission was reported
in a survey conducted at the Owens Valley Radio Observatory 
(\citealt{zl87}).
\citet{bwhz94} detected 5 GHz radio continuum emission 
in the direction of \iras 18576 in their Galactic Plane Survey 
using the Very Large Array (VLA) and catalogued the source as 
GRSR5 37.278$-$0.226.
Based on the positional coincidence, they determined
the matching probability of the radio source to
\iras 18576 to be 98.4\% \citep{wbh91, bwhz94},
suggesting the source to be a PN based on the \iras flux density 
ratios.
In the NRAO VLA Sky Survey (NVSS; \citealt{ckt99}),
1.4 GHz radio emission was measured despite the confused 
background.

\citet{hvk00} observed \iras 18576 with 
the Short Wavelength Spectrometer (SWS; \citealt{degraauw96}) 
on board the
Infrared Space Observatory ({\it ISO\/}; \citealt{kessler96}).
The spectrum shows an IR excess peaking near $27 \um$ with 
unidentified IR (UIR) features at 3.3, 6.2, 7.7, 8.6, and 11.3 $\um$, 
which are often attributed to polycyclic aromatic hydrocarbons 
(PAHs; \citealt{ahs99} and references therein).
The \iso detection of carbon-rich (C$-$rich) material, together 
with the previous \orich suggestion by the \iras LRS,
makes the source both C/O$-$rich object. 
The authors provided several possibilities for the
nature of \iras 18576, which include multiple central
sources (\ion{H}{2} regions or PNs) and a source with
multiple shells with differing composition.
Table 1 summarizes previous photometric observations 
of \iras 18576.

\section{Observations and Results}

\subsection{Mid-IR Imaging}

We obtained images of \iras 18576 using the University 
of Arizona/Smithsonian Astrophysical Observatory Mid-IR Array 
Camera (MIRAC3, \citealt{hoffmann98}), at the NASA Infrared 
Telescope Facility (IRTF) on 1999 June 7 under photometric 
conditions.
The array is a Boeing HF-16 arsenic-doped silicon blocked-impurity-band
hybrid array and has a $128 \times 128$ pixel format.  
The pixel scale was set to $0\arcsec.33$ ($42\arcsec \times 42\arcsec$ 
field of view), which ensures a Nyquist sampling of the 
diffraction-limited PSF of the telescope. 
The object was observed with $10\%$ ($= \Delta \lambda/\lambda$) 
bandwidth at 10.3 and $18.0\um$.
For flux and point-spread-function (PSF) calibration,
we observed $\alpha$ Her (a CGS3 standard; \citealt{cohen95})
before and after the object to check for variations in the PSF. 
Measured PSF size (FWHM) was $1\arcsec.05 \pm 0\arcsec.06$ 
for $10.3 \um$ and $1\arcsec.53 \pm 0\arcsec.05$ for $18.0 \um$.
With an east-west telescope nod throw and a north-south secondary
chop throw, we mosaiced a final image of size about $25\arcsec$ across.
Flat-fielding and bad-pixel-masking were applied to 
sky-subtracted, co-added images before mosaicing.
Each individual image was subdivided by $4 \times 4$ 
during mosaicing to make the pixel scale of $0\arcsec.0825$/pix 
for accurate registration.
The total integration times of the final images
are 720 and 400 sec respectively at 10.3 and $18.0 \um$,
resulting in 1 $\sigma$ rms noise of 14 and 94 mJy arcsec$^{-2}$.
Absolute flux calibration errors are estimated to be approximately $10\%$.
A full description of the nod-chop procedure,
data reduction, and flux calibration processes can be found
in \citet{meixner99}.

Figure 1 shows the reduced images of \iras 18576 at 10.3 and $18.0 \um$.
The dust nebula appears roughly circular with about $7\arcsec$ 
diameter and there are local emission peaks in the inner region 
of the nebula. 
At $10.3 \um$ (Figure 1a), three emission peaks are clearly seen and 
are aligned to form a straight line whose position angle (east 
from north) is $-3.5^{\circ}$.
We shall refer to the peaks as northern, central, and southern
peaks according to their relative positions.
The central peak, located near the geometrical center of the nebula,
is unresolved and very likely represents the central star of the system.
The northern and southern peaks are $2\arcsec$ to $2\arcsec.5$
away from the central peak and more extended than the central peak.
The arc-like elongation of the northern and southern peaks seems 
to encircle the central peak.
At $18.0 \um$ (Figure 1b), the morphological characteristics are 
identical to those at $10.3 \um$ except that the central peak is 
less pronounced at $18.0 \um$.
The total specific flux of the nebula is 16.7 Jy and 205 Jy 
respectively at 10.3 and $18.0 \um$:
the northern peak accounts for $\sim 70\%$ of the total emission
while the rest is shared by the southern peak ($\sim 25\%$) and
the central peak ($\sim 5\%$) in both wavebands.
Emission from the central peak would impose an upper 
limit for the direct stellar emission,
and we estimated the specific flux from the central star 
at each wavelength following the iterative method 
described in \citet{hawkins95}.

Figure 2 shows the temperature and optical depth maps, which are 
derived from the source images by following the method described 
by \citet{dayal98}.
Temperature  (Figure 2a) is almost constant over the entire 
extent of the nebula: $115 - 135$ K with an average of 122 K.
There is a single temperature peak coinciding with the
central emission peak and temperature falls off as the
distance from the central peak increases.
This indicates that dust grains in the central peak have
higher temperature than dust grains in the surrounding
nebula (northern and southern peaks).
Thus, it is reasonable to assume that there is a single 
energy source located in the central emission peak and 
that the surrounding material is heated by radiation from 
the central source.
This supposition agrees with the behavior of the relative 
strengths of the emission peaks in the nebula.
While the relative emission strength of the northern and 
southern peaks remains constant,
that of the central peak does change at two wavelengths.
The observed decrease in the relative emission strength of
the central peak at the longer wavelength is a natural
consequence at the Rayleigh-Jeans tail of the
blackbody curve with high photospheric/dust temperature, 
whereas the blackbody 
curve with typical dust grain temperature ($\ale 200$ K)
generally peaks at mid-IR regions and would not lead to
a change in the relative emission strength.
The optical depth maps of the dust shell are both similar
at $10.3 \um$ (Figure 2b) and $18.0 \um$.
They indicate that the nebula is optically thin
($\tau_{\rm max} = 0.058$ and 0.028 respectively at 10.3 and $18.0 \um$)
and the northern and southern emission peaks coincide with
the regions in which the optical depth is high.
This is corroborated by the fact that evidence for
the central object can be seen both at 10.3 and $18.0 \um$.
Therefore, we interpret the mid-IR morphology of \iras 18576 
as limb-brightened edges (northern and southern peaks) of an 
optically-thin edge-on dust torus surrounding the central 
star, as seen in, for example, HD 168625 (\iras 18184$-$1623; 
\citealt{rh98, meixner99}) and $\eta$ Car \citep{polomski99}.
We define the inner radius of the dust torus ($\rdust$) to be 
$2\arcsec.4$, which is the distance from the central star
to the northern and southern peaks.
In Table 2, we summarize quantities of the CDS in \iras 18576
measured and derived from the mid-IR images.

\subsection{Near-IR Photometry}

To check the variability of \iras 18576,
we obtained {\it JHK'} photometry of the object 
with a 40 inch telescope equipped with the Near-Infrared 
IMager (NIRIM; \citealt{myol99}) at Mt. Laguna 
Observatory\footnote{Mt. Laguna Observatory is jointly 
operated by San Diego State University and University 
of Illinois at Urbana-Champaign.} 
on 1999 June 12 and Nov 14 and 2000 August 5 under clear skies.
For flux calibration, we used ELIAS IR standard \citep{elias82} 
and UKIRT faint IR standard stars \citep{casali92}.
The standard stars were observed at similar airmasses as
the source or at low ($\sim 1$) and high ($\age 2$) airmasses 
to perform an airmass correction.
Data were taken by shifting the telescope with a 9-point
dithering pattern.
Each exposure was flat-fielded to eliminate large 
pixel-to-pixel sensitivity variations in the detector 
array and was then sky-subtracted before being co-added
into a single frame.
Flats were created for each waveband by exposing the
twilight sky. 
Sky emission maps were constructed by taking a median 
of all dithered frames after unusually low and high pixels 
had been masked out.
Our reduction method generally follows the method described
in \cite{mclean96}.
Table 3 shows {\it JHK'} photometric data for \iras 18576,
in which all magnitudes have been converted to the ELIAS 
(CIT) magnitudes \citep{elias82} for comparison.

It appears that near-IR brightness of \iras 18576 is 
in decline in the past decade with the most recent
observation (2000 August 5) marking the dimmest ever.
Because we did not detect a similar brightness variation in 
other objects observed on the same nights during the NIRIM 
observations, we eliminate the possibility of any
systematic and/or instrumental effects.
Therefore, it is likely that \iras 18576 varies its near-IR 
brightnesses by a few tenth of a magnitude over a couple of 
months and about 0.5 magnitudes over a decade.

\section{Radiative Transfer Modeling}

\subsection{2-D Dust Radiative Transfer}

To model IR continuum emission and mid-IR morphologies,
we have performed model calculations for an axisymmetric CDS
by using a code which solves the radiative transfer equation
in a fully two-dimensional grid.
This code treats dust absorption, reemission, and (isotropic) 
scattering in an equatorially-enhanced dust distribution 
(i.e., a presumed radial and latitudinal density variation)
with an arbitrary optical depth for a given set of dust species
of a single grain size
(\citealt{meixner97,skinner97} for more description of the code).
The equatorially-enhanced dust distribution in the model CDS
assumes a two-phased mass loss process, in which the central star
changes its mode of mass loss from spherically symmetric 
to axially symmetric with a preferential mass loss towards 
equatorial directions near the end of the entire mass loss phase.
As a result, the pole-to-equator density ratio at the inner 
shell boundary can be less than unity.
The code requires input parameters that can be categorized
into three types: stellar, shell, and dust parameters
(Table 3).
Stellar parameters (the effective temperature, $\teff$, 
radius, $\rstar$, and distance) determine the available 
energy flux for dust heating.
Shell parameters (inner and outer shell radii, $r_{\rm d,~min}$ 
and $r_{\rm d,~max}$, optical depth along the polar direction 
at some reference wavelength, $\tau_{\lambda}$,
and other geometrical parameters) specify the dust shell 
(both physically and morphologically) 
for radiative transfer calculations.
Dust parameters (grain radius and absorption and scattering
coefficients, $Q_{\rm abs}$ and $Q_{\rm sca}$) determine 
opacity for the shell material.

The use of this dust radiation transfer code is appropriate 
because direct stellar radiation is estimated to be the 
dominant heating source for the circumstellar dust grains. 
The total far-infrared flux (observed specific flux 
integrated beyond $20 \um$) of \iras 18576 is 
$\sim 3.9 \times 10^{-11}$ W$^{2}$ m$^{-2}$.
However, we estimate the amount of far-infrared flux expected 
from dust grains heated by Ly $\alpha$ flux (i.e., line emission)
to be $\sim 8.7 \times 10^{-13}$ W$^{2}$ m$^{-2}$,
following the the method described by \citet{zijlstra89}.
Thus, we can safely neglect dust heating due to line emission.

When available, stellar and dust parameters are taken
from previous observations in the literature. 
Otherwise, we determine these parameters through
iterative model calculations seeking the best-fit to the
spectral energy distribution (SED).
Once we achieve the best-fit to the SED, we further 
iterate calculations with different sets of shell parameters
and inclination angle\footnote{The inclination angle is 
the sharp angle between the line 
of sight and the polar axis.} ($\theta_{\rm incl}$) 
seeking the best-fit
to the SED as well as the mid-IR morphology, which tends 
to be strongly influenced by the geometric parameters.
For example, limb-brightened peaks of a dust torus would not 
be resolved with a too high inner shell optical depth and
a single shell structure can yield different morphologies 
depending on the inclination angle.
The overall shape of the nebula would be ring-like
if $\theta_{\rm incl} \sim 0$ and 
would be flattened in the direction of the toroidal axis
if $\theta_{\rm incl} \sim 90$.
The final best-fit model was achieved, following the guidelines 
explained above, after $\sim 200$ model calculations
shifting carefully through the parameter space.
Because our code does not handle dust stratifications 
at the moment, the final best-fit model was constructed 
by combining separate best-fit models for each of the
\orich and \crich components of the dust shell,
assuming the cavity inside the axisymmetric \orich shell 
is filled with spherically symmetric \crich material.

\subsection{Modeling Results}

Figure 3 shows the best-fit SED of our model calculations.
The best-fit SED (thin solid line) consists of possibly 
stratified shells of differing composition:
the outer \orich shell (thin dashed line) and the inner 
\crich shell (thin dash-dotted line).
For comparison, the \iso spectrum (thick solid line), 
the \iras LRS\footnote{\iras
LRS spectrum does not have an absolute flux calibration
and is shown primarily to show the shape.
The best-fit was determined by comparing model fluxes with
the \iso spectrum and other photometric data.}
(thick dashed line), and other photometric data (symbols)
are shown. 
Table 3 lists input and derived quantities for the 
best-fit model.

We also show the best-fit synthesized images at 10.3 and 
$18.0 \um$ (Figure 4).
The final synthesized images were constructed by 
adding the observed standard star images to
the \orich shell model images because the code does not 
produce images with a resolved central star.
The model images were convolved with a Gaussian 
profile whose FWHM is equal to the observed FWHM at each 
waveband before the central star was added.
The two emission peaks arise from equatorially-enhanced 
(the pole-to-equator density ratio $= 0.5$)
\orich dust distribution viewed with $45^{\circ}$ 
inclination angle.
The resulting morphologies are very sensitive to the 
inclination angle when the shell is optically thin and 
would completely be altered if the inclination angle is
changed by $\sim 10\%$.
The axis of the dust torus is rotated in the plane of the
sky at a position angle of $86.5^{\circ}$ East of North
and
the near side of the dust torus is to the east from the 
center.
The slight asymmetry seen in the shape of each 
emission peak is due to self-extinction arising
from the polar density variation. 

\section{Discussions}

\subsection{The Central Star and its Distance}

Because we have no prior knowledge about the central star, 
we must iteratively derive $\teff$ and $\rstar$ from
radiative transfer calculations.
Consider an optically thin, isothermal (at $\tdust$) dust shell.
When dust grains of a uniform grain size are in
thermal equilibrium with the ambient stellar radiation field,
we can relate the luminosity of the star ($L_{*}$),
shell radius ($\rdust$), and dust temperature 
by equating the rate of energy absorption and reemission as
\begin{equation}
 \tdust \propto
 \left( \frac{\lstar}{\rdust^{2}} \right)^{\frac{1}{n+4}}.
 \label{dusttemp1}
\end{equation}
Here, we have assumed a power-law dust grain emissivity
($\qabs \propto \nu^{n}$) and the constant of proportionality
is solely determined by the dust grain properties.
Because both $\lstar$ and $\rdust$ are observable quantities
modulo distance, we can eliminate the distance dependence
in  eq.(\ref{dusttemp1}) and rewrite it as
\begin{equation}
 \tdust \propto \left( \frac{\fstar}{\rdust'^{2}} \right)^{\frac{1}{n+4}},
 \label{dusttemp2}
\end{equation}
in which $\fstar$ is the flux of the source
and $\rdust'$ is the angular radius of the dust shell in 
arcsec.
The dust temperature effectively defines the 
location of the IR excess peak in the SED. 
Therefore,
eq.(\ref{dusttemp2}) indicates that we can fix $\fstar$
by fitting the IR peak location (i.e., $\tdust$) by iterating on
$\fstar$ and dust species in radiative transfer calculations
because $\rdust'$ is observationally determined.
It is also possible to constrain $\fstar$ by using
the observed IR flux of the source ($\fstarobs$).
Photometric data obtained by \iso \citep{hvk00} and other
near-IR observations \citep{gmpp97} yielded
$\fstarobs \sim 2.3 \times 10^{-7}$ erg s$^{-1}$ cm$^{-2}$.
The near-IR part of the \iso spectrum \citep{hvk00}
suggests the presence of PAHs.
However, solely \crich models yielded a 
warmer dust temperature than can be produced by the 
detached dust shell of $2\arcsec.4$ angular radius
due to generally higher opacity of carbonaceous material.
Models with a mixture of amorphous carbon and astronomical 
silicates produced too high dust temperature as well.
Our iterative calculations concluded 
that $\fstar \sim 4 \times \fstarobs$ with astronomical
silicates would produce the best-fit mid-IR SED to observations.
This means that the \orich dust shell processes about 
$25 \%$ of the stellar flux,
and it is consistent with our finding that the dust shell
of \iras 18576 is optically thin.

To recover $\teff$ and $\rstar$ from the derived $\fstar$, 
we need to determine the distance to \iras 18576.
Previous CO observations do not provide any kinematic information
of the source \citep{zd86} probably because of the 
confused background reported by \citet{ckt99}.
Recent CO observations failed to obtain any reliable
kinematic data due to the lack of clear background
in the crowded radio field around \iras 18576 
(D. Fong 1999, priv. comm.).
Therefore, we estimated the distance to \iras 18576 by
determining the total visual extinction 
($A_{\rm V}$) that is required to make our model near-IR 
SED fit the observed one.
The near-IR colors of \iras 18576 are very red.
Because the dust shell is assumed to process only about
$25\%$ of the stellar radiation, it is reasonable to
explain the red near-IR colors by a severe interstellar
extinction.
Our best fit model indicated that $A_{\rm V}$ would have to
be about 28 to account for the extinction
assuming a typical total-to-selective extinction ratio 
($R_{\rm V}$) of 3.1.
Such a high $A_{\rm V}$ value may not be surprising 
because of the Galactic coordinates of IRAS 18576,
$(l,b) = (37.3^{\circ}, -0.3^{\circ})$:
the line of sight lies in the Galactic plane
grazing the outskirts of the bulge. 
According to the full-sky dust maps compiled from 
{\it COBE\/}/{\it DIRBE\/} and \iras maps
\citep{schlegel98}, the Galactic selective extinction, $E(B-V)$,
along the line of sight to \iras 18576 is estimated to be 
$18.25$, which equals to the Galactic
total visual extinction of 56.58.
If we assume the Parenago type exponential distance scaling
of $A_{\rm V}$ (\citealt{parenago40}),
$A_{\rm V} \sim 28$ can be achieved if \iras 18576 is 
located around 10 kpc away.
This estimate is uncertain because the assumed
distance scaling law of $A_{\rm V}$ does not account for
local clumping of reddening material and may locally have 
a drastically different form.

Alternatively, the distance to \iras 18576 can be estimated 
by iteratively searching for a high enough $\lstar$ to heat 
the optically thin dust shell of \iras 18576 (Figure 2b) 
to the observed temperature ($\tdust = 115 - 135$ K)
because the distance directly scales the luminosity of the 
object.
The optical depth maps have yielded very low $\tau$ values
($\tau_{\rm max} = 0.058$ and 0.028 at 10.3 and $18.0 \um$,
respectively).
With these optical depth, \iras 18576 would have to
be located about 25 kpc, which yields an unrealistically
high luminosity for a star and brings the object close to 
the edge of the Galaxy in the direction of the object.
Thus, we kept increasing the optical depth of the dust shell
until the toroidal structure would be unresolved
while iteratively adjusting the distance.
After further iteration, we concluded that 
$10 \pm 3$ kpc would scale $\lstar$ high enough 
so that model calculations would yield the dust peak in the 
observed SED,
using the highest optical depth possible to 
reproduce the resolved dust toroidal structure in our 
models (Table 3 and Figure 4).
Considering the above two independent means to estimate
the distance, 
we adopted 10 kpc to be the distance to \iras 18576
because there is no alternate source of data.
With the adopted distance of 10 kpc, $\lstar$ is
$1.1 \times 10^{40}$ ergs s$^{-1} (= 10^{6.4} \lsun)$.

We can then constrain $\teff$ from the position of the 
redward slope of the stellar emission peak with respect 
to the observed SED.
A lower $\teff$ would shift the stellar peak to the red
and would effectively increase the stellar emission in the
wavelengths of our interest while leaving the dust peak
virtually unchanged.
Thus, the lower limit for $\teff$ can be obtained 
because any lower $\teff$ than the limit would yield
too much stellar flux than observed.
We used the $10.3\um$ emission in constraining $\teff$
because this wavelength range seemed to retain the most 
pristine stellar emission characteristic free from 
extinction and/or dust feature emission.
Further iteration yielded the lower $\teff$ limit of
$\sim 9000$ K, which corresponds to the upper $\rstar$ 
limit of $\sim 700 \rsun$.
However, it is more difficult to constrain the upper $\teff$
limit defining the possible range for $\teff$
because the emission properties in the optical to 
near-IR depend more heavily on the distance to the object 
(i.e., $A_{\rm V}$) and 
the location of the \crich shell that is
considered to be situated inside the \orich dust 
shell (see below).
Nevertheless, iterative calculations suggested that 
$\teff \sim 15000 \pm 6000$ K would be a likely
range of the effective temperature.
The model calculations with this $\teff$ range requires
$A_{\rm V} = 24 \sim 32$ to be consistent with the
observed data.

\subsection{Dust Shell Composition and Structure}

The presence of silicates in \iras 18576 was originally 
suspected from the apparent absorption feature at $10 \um$ 
\citep{zd86,vc89} seen in the LRS.
This $10 \um$ dip, however, is a false depression 
caused by the strong neighboring PAH features at 8.6 and 11.3 $\um$,
because the CDS is observed to be optically thin enough that 
the central star is visible even at the mid-IR (Figure 1).
This makes it difficult to identify the silicate species 
from the shape of the $9 - 11 \um$ silicate features
(\citealt{speck98} and references therein).
The featureless far-IR portion of the spectrum also makes it
difficult to determine if there exists a particular 
crystalline silicate species.
Therefore, we used optical constants for astronomical silicates
\citep{ld93} which empirically simulate a mixture of 
different types of silicates.
With iteratively obtained $\fstar \sim 4 \times \fstarobs$,
the model SED fits the overall shape of the \iso SED very 
well at the long wavelength regions ($\age 18 \um$) and at 
the $10 \um$ trough (Figure 3, thin dashed lines).

The model SED arising only from the \orich shell shows 
emission ``deficits'' in the near-IR ($\ale 10 \um$) and at 
the blueward shoulder of the IR peak ($10 - 18 \um$).
To account for the presence of the observed \crich material,
we considered an inner \crich (PAHs) region, which is
surrounded by the \orich shell of angular radius $2\arcsec.4$.
\crich material is expected to be closer to the central 
star because of the presumed ionization state of PAHs:
the comparable strengths of 7.7 and 11.2 $\um$ emission peaks
suggest that PAHs are marginally ionized \citep{molster96}.
The inner \crich shell adequately compensates 
the emission deficits (Figure 3, thin dot-dashed lines).
The \crich shell does not reproduce each emission line of 
the UIR features because we have used optical constants 
of amorphous carbon grains of $0.002 \um$
to substitute for the yet unavailable optical constants of PAHs.
This dust grain size is comparable to
the major PAH cluster size of $300 - 400$ C$-$atoms 
\citep{omont86} whose presence is typically observed
through $5 - 10 \um$ and $11 - 15 \um$ 
emission plateaus \citep{beintema96}.
Due to lack of spatial information in the near-IR,
we determined the inner \crich shell radius through
model iterations.

The deficit at the longer wavelengths ($15 - 20 \um$), on the 
other hand, seems to be excessive because very little PAH 
emission is expected in this wavelength region.
To boost IR emission from 15 to $20 \um$, we 
included amorphous aluminum oxide (Al$_{2}$O$_{3}$), whose 
$\qabs$ has a local maximum around $13 \um$, according to 
the mass abundance ratio consistent with the cosmic abundance
($\sim 96\%$ amorphous silicates and $\sim 4\%$ aluminum oxides).
The inclusion of aluminum oxide grains is not entirely 
{\it ad hoc}; they are considered to be the nucleation cores
in the heterogeneous dust nucleation theory (\citealt{speck98}
and references therein)
and the \iso spectrum shows a number of possible narrow peaks 
in the $11 - 13 \um$ plateau (especially 11.9 and 12.8 $\um$),
which may be due to aluminum oxides \citep{begemann97}.
Although we did not attempt model calculations with any 
specific silicate species, the presence of an emission 
plateau near $34 \um$, which our model SED fails to 
reproduce (due to the use of amorphous silicates), suggests the
presence of crystalline olivines (forsterites), which 
have an emission peak in that region.

Our best-fit model agrees with the general behavior of the 
observed SED with the 10 $\um$ trough adequately reproduced.
This method obviously does not conserve flux at the
interface between the two shells in the sense that the 
\orich shell receives stellar flux which should have 
been reduced by the presence of the \crich shell.
However, the very optically thin \crich shell is estimated
to require only about $0.5\%$ of the total flux of the 
central star.
Therefore, considering that the \orich shell itself processes 
only $\sim 25\%$ of the total stellar flux, we conclude
that our results would still hold even if calculations 
were done in a strictly flux conserving manner.

\subsection{Nature of IRAS 18576+0341}

The derived stellar parameters put the star above the
empirical luminosity limit (Humphreys-Davidson limit;
\citealt{humphreys79})
in the Hertzsprung-Russell diagram.
\iras 18576 would then likely be a luminous blue variable 
(LBV) or a red super giant (RSG).
The derived luminosity of \iras 18576 is unusually high 
($\lstar = 10^{6.4} \lsun$) for a RSG, however,
there exist LBVs whose $\lstar$ is comparable to that of 
\iras 18576
(e.g., $\eta$ Car and AG Car, \citealt{humphreys94} and 
references therein).
Although the expected temperature range for the object 
is rather low ($\teff = 15000 \pm 6000$ K) for a typical 
LBV,
it is still higher than that of the coolest LBVs.
The fact that $\teff$ would not be
lower than 9000 K seems to favor the LBV interpretation.
One of the coolest LBV candidates, HD 168625, 
was found to have $\teff$ in the range of $12000 - 15000$ K 
\citep{nota96}.
Interestingly, this cool LBV is also found to be associated 
with an extended dust shell with the equatorial enhancement
\citep{rh98, meixner99}.

There are a number of observational similarities
between \iras 18576 and other LBVs.
The mid-IR morphology of LBVs is characterized especially 
by their signature limb-brightened lobes with
a separate central source
(e.g., $\eta$ Car, \citealt{polomski99}; 
AG Car, \citealt{trams96}; 
HD 168625, \citealt{rh98,meixner99}),
which are very similar to what we have seen in \iras 18576
(Figure 1).
Morphologically, it is very likely that the central 
emission peak represents the one and only energy source
in the system and illuminates the surrounding lobes.
The limb-brightened lobes strongly suggest the presence of
an equatorially enhanced dust distribution in the CDS,
and it is consistent with what our model calculations assume.
This axisymmetric geometry has been thought to be 
a common signature of the LBV nebulae caused by 
a sharp pole-to-equator density contrast \citep{nota95}.
Although the origin of the equatorial 
enhancement is not yet fully understood,
a number of possible mechanisms have been suggested.
For example,
\citet{langer99} considered rotating stars near their
Eddington limit and showed hydrodynamically that
highly non-spherical mass loss would occur.
On the other hand, \citet{damineli00} have 
spectroscopically confirmed the binarity of $\eta$ Car,
and it has also been hydrodynamically shown that 
such binary systems can generate an equatorially-enhanced
material distribution around the central system in
the context of the PN formation \citep{mm99}.
Whatever the true scenario may be, the observed 
limb-brightened mid-IR lobes clearly showed the 
non-sphericity of the system, and any dynamical 
models should be able to reproduce such geometry.

Spectral characteristics of \iras 18576 are also similar
to those of other LBVs.
Large IR excesses due to dust grains have been observed
in LBVs, and 
the observed mid-IR flux levels of \iras 18576 are 
comparable to other LBVs:
it is approximately a few times brighter than typical
LBVs \citep{lamers96} but a couple of orders of magnitudes
dimmer than $\eta$ Car \citep{morris99}.
Both amorphous and crystalline silicates have been 
identified in the LBV dust shells \citep{waters97}.
The presence of \crich material in a dominantly
\orich shell has already been seen in other LBVs
(e.g., \citealt{voors99}).
Moreover, \iras 18576 seems to be varying its brightness
with the timescale  (months to years) and the widths of 
magnitude (about $0.1 - 0.5$) consistent 
with what have been observed in LBVs.

The observed and derived physical quantities of the dust 
shell are also consistent with LBV characteristics.
The maximum extent of the observed mid-IR dust nebula
($\age 5 \sigma$ detection) is about $10\arcsec$, which 
equals to $\sim 0.5$ pc at 10 kpc.
Considering the fact that the mid-IR observations probe 
only the innermost regions of the dust shell,
this value is reasonable for the inner radius of a typical
LBV nebula (e.g., \citealt{nota95, chu99}).
The total dust mass in the \iras 18576 dust torus is
estimated to be $\sim 0.1 \msun$.
This estimate depends on the outer radius of the dust torus, 
which is assumed to be 6 times larger than the inner radius.
The estimated dust mass is nearly a factor of 10 higher than
typical LBV dust mass quantities \citep{nota95} but is
comparable to the estimated dust mass of a similar 
temperature ($100 \pm 10$ K) component in 
$\eta$ Car \citep{morris99}.
The total mass lost by the star and the rate of mass loss 
are estimated to be about $75 \msun$ and 
$6.8 \times 10^{-6} \msun$ yr$^{-1}$ using the 
canonical gas-to-dust ratio of 100
and an assumed expansion velocity of $50$ km s$^{-1}$ 
(e.g., \citealt{nota95}).
Both of these values depend on the outer radius of the entire
mass loss shell, which is not constrained by observations and 
is assumed to be 50 times as the inner shell radius.

Based on the mid-IR morphology and IR photometric
and spectroscopic characteristics, and results from 
radiative transfer calculations,
\iras 18576 is likely a very luminous star surrounded by
an optically thin, possibly stratified C/O$-$rich dust shell
located at about 10 kpc in the direction in which 
very severe interstellar extinction can occur.
Physical parameters of the \iras 18576 system derived from 
radiative transfer calculations suggest that \iras 18576
is probably a LBV candidate.
This conclusion derived from the fact that the derived
luminosity for \iras 18576 was so high and that other
evidence seemed to be consistent with our conclusion.
However, given the uncertainties involved in the distance 
determination, it is also possible that \iras 18576 is 
a RSG.
Therefore, it is necessary to observationally uncover 
the physical characteristics of the central star
to determine the exact evolutionary status of \iras 18576.
Because \iras 18576 is not visible in the optical
due to heavy extinction, we need near-IR spectroscopic
information to constrain the physical parameters of 
the central star.
We also need to continue obtaining near-IR photometry
to  better characterize the variability of the source.
We are planning such follow-up observations.

Regardless of the exact evolutionary status,
\iras 18576 seems highly likely to be an evolved, massive 
post-main-sequence star.
The axisymmetry of the circumstellar shells of evolved
massive stars has long been indirectly suggested from 
the strongly non-spherical shapes of optical nebulosities 
associated with LBVs (e.g., \citealt{nota95}
and references therein).
Our discovery of the dust shell around \iras 18576 is one 
of the few direct evidence for the equatorial density 
enhancement in the circumstellar environment of massive 
post-main-sequence stars
(e.g., $\eta$ Car, \citealt{polomski99}; 
AG Car, \citealt{trams96}; 
HD 168625, \citealt{rh98,meixner99}).
For low to intermediate mass stars, it is well established 
that the mode of mass loss is highly axisymmetric
(e.g., \citealt{ueta00}).
It now appears that an axisymmetric mass loss also operates 
for massive stars, and therefore,
stellar wind scenarios for massive stars would be able to
produce the observed pole-to-equator density contrast
and generate axisymmetric CDSs.

\section{Conclusions}

We have discovered an extended mid-IR emission around \iras 18576
and presented the mid-IR images at 10.3 and 18.0 $\um$.
The two emission peaks, which are concentrically elongated and 
symmetrically oriented on the opposite sides of the central peak, 
are typically the results from an optically thin, limb-brightened 
edges of the dust torus viewed rather edge-on.
We also present new near-IR photometry of the object which
demonstrates its variability.
Although the source lies in the region of the sky in which
a severe Galactic extinction has been observed,
we have constructed a model using all of the available information.
The derived physical quantities for the central star and dust 
shell and variability of the object
suggest that \iras 18576 is a possible LBV candidate with 
$\lstar = 10^{6.4} \lsun$ and $\teff = 15000 \pm 9000$ K.
The uncertainties involved in the distance determination
leave possibilities that the object could be a RSG,
and further observational work is needed to determine
the exact evolutionary state of the star.
While the main dust shell is composed of silicate dust grains, 
an additional \crich (composed of PAHs) shell likely exists
inside of the \orich shell.
The axisymmetric nature of the dust toroid is characterized
by the calculated pole-to-equator density ratio of 0.5.
While the origin of the equatorial enhancement is still open 
for further discussion,
the axisymmetry seems to be a common feature in mass loss 
from massive stars as in the case for low to intermediate 
stars.

\acknowledgments

Ueta and Meixner are supported by NSF CAREER Award AST-9733697.
We would like to thank 
A. K. Speck for enlightening discussions on dust properties,
B. J. Hrivnak, K. Volk, and S. Kwok for providing their ISO spectra,
and 
D. Fong for performing the CO observation.
We are also grateful to an anonymous referee for many 
useful comments and suggestions.
This research has made use of the Jena - St. Petersburg 
Database of Optical Constants (JPDOC;
http://www.astro.uni-jena.de/Users/database/f-dbase.html)
operated by the Astrophysical Institute of the Friedrich 
Schiller University, Jena, Germany, and the Astronomical 
Institute of the St. Petersburg University, Russia,
and 
the SIMBAD database, operated at Centre de Donn{\'e}es astronomiques, 
Strasburg, France.

\clearpage
\figcaption{Observed grayscale images of \iras 18576+0341 at 
$10.3 \um$ (a; top) and $18.0 \um$ (b; bottom) with north 
being up and east being to the left.
The tick marks show relative offsets in arcseconds.
Contours are spaced by $10\%$ of the peak intensity 
and their colors are inverted for the ease of viewing.
The lowest contour is equivalent of $4.3 \sigma$ 
and $6.7 \sigma$ level of emission respectively at 
10.3 and 18.0 $\um$.
Local peak intensities are, from northern to southern peak,
0.57, 0.49, and 0.38 Jy arcsec$^{-2}$ at 10.3 $\um$
and 6.3, 3.9, and 4.4 Jy arcsec$^{-2}$ at 18.0 $\um$.
The PSF size (FWHM) is indicated by a black circle in each frame
at the lower left: 1\arcsec.05 at $10.3 \um$ and 1\arcsec.53
at $18.0 \um$.}

\figcaption{Temperature map (a; top) and optical depth map
at $10.3 \um$ (b; bottom) of \iras 18576 following the 
display convention of Figure 1.
(a) Temperature contours go from 135 to 115 K with an interval of
2.5 K.
The spurious peaks around the nebula are artifacts occurred
during the derivation.
(b) Optical depth contour interval is 10\% of the peak value,
0.058.}

\figcaption{The spectral energy distribution of the best-fit model 
(9000 K, $700 \rstar$, 10 kpc): 
thin solid, dashed, and dot-dashed lines respectively show
the total SED, contribution from the \orich shell, and contribution
from the \crich shell with ISM extinction (bottom lines) and
without extinction (top lines).
Observational data are indicated by thick lines 
(solid $-$ \iso; Hrivnak et al.\/ 2000,
dashed $-$ \iras LRS; Volk and Cohen 1989)
and symbols 
(circles $-$ \iras photometry; \iras Explanatory Suppliment 1988,
asterisks $-$ AFGL photometry; Price \& Murdock 1983,
crosses $-$ near-IR photometry; Garc\'{\i}a-Lario et al.\/1997, 
stars $-$ MSX photometry; Egan et al.\/1999,
squares; this observation).}

\figcaption{Model Images of \iras 18576+0341 at $10.3 \um$ (a; top)
and $18.0 \um$ (b; bottom)  following the 
display convention of Figure 1.
Contours are spaced by $10\%$ of the peak intensity.
Peak intensities are 0.40 Jy arcsec$^{-2}$ (at 10.3 $\um$) 
and 8.3 Jy arcsec$^{-2}$ (at 18.0 $\um$).
The central star (the observed PSF) was scaled to reflect 
the observed northern-to-central peak ratio at each wavelength.}

\clearpage
\begin{deluxetable}{lccccccccccccc}
\tabletypesize{\scriptsize}
\tablecolumns{14} 
\tablewidth{0pc} 
\tablecaption{Previously Obtained Photometry on \iras 18576$+$0341} 
\tablehead{ 
\colhead{RA (2000)} &
\colhead{DEC (2000)} &
\colhead{$11.0 \um$\tablenotemark{a}} &
\colhead{$12.0 \um$\tablenotemark{b}} &
\colhead{$19.8 \um$\tablenotemark{a}} &
\colhead{$25.0 \um$\tablenotemark{b}} &
\colhead{$27.4 \um$\tablenotemark{a}} &
\colhead{$60.0 \um$\tablenotemark{b}} &
\colhead{$100.0 \um$\tablenotemark{b}} &
\colhead{5 GHz\tablenotemark{c}} &
\colhead{1.4 GHz\tablenotemark{d}} \\
\colhead{} &
\colhead{} &
\colhead{(Jy)} &
\colhead{(Jy)} &
\colhead{(Jy)} &
\colhead{(Jy)} &
\colhead{(Jy)} &
\colhead{(Jy)} &
\colhead{(Jy)} &
\colhead{(mJy)} &
\colhead{(mJy)} }
\startdata 
19:00:11.2 & +03:45:46 & 81.4 & 58.48 & 356.9 & 424.2 & 649.4 & 
274.5 & $< 1661$ & 78.1 & $8.1\pm0.7$ \\
\enddata \\
\tablerefs{a: AFGL survey \citep{pm83}, b: \iras \citep{iras},
c: GPS survey \citep{bwhz94}, d: NVSS survey \citep{ckt99}}
\tablecomments{See Table 4 for the near-infrared photometric data
and Figure 3 for \iras LRS and \iso spectra.
\iras $100.0 \um$ data is an upper limit.}
\end{deluxetable}

\clearpage
\begin{deluxetable}{lcc}
\tablecolumns{3} 
\tablewidth{0pc} 
\tablecaption{Measured and Derived Quantities of the Dust Shell around \iras 18576$+$0341} 
\tablehead{ 
\colhead{Quantity} & 
\colhead{10.3 $\um$}   & 
\colhead{18.0 $\um$}} 
\startdata 
Flux Density (Jy) & 16.7     & 205 \\
Diameter (FWHM) & 6\arcsec.2 & 6\arcsec.9 \\
Toroidal Shell Radius & \multicolumn{2}{c}{2\arcsec.4}  \\
Dust Temperature & \multicolumn{2}{c}{115 $\sim$ 135 K} \\
Optical Depth    & $\ale$ 0.058 & $\ale$ 0.028 \\
\enddata 
\end{deluxetable}

\clearpage
\begin{deluxetable}{ccc}
\tabletypesize{\scriptsize}
\tablecolumns{3} 
\tablewidth{0pc} 
\tablecaption{Input and Derived Model Quantities} 
\tablehead{\multicolumn{3}{c}{Stellar Parameters}}
\startdata 
$\lstar$ & \multicolumn{2}{c}{$10^{6.4} \lsun$} \\
$\teff$  & \multicolumn{2}{c}{$15000 \pm 6000$ K} \\
$\rstar$ & \multicolumn{2}{c}{$120 - 700 \rsun$} \\
$d$      & \multicolumn{2}{c}{$10 \pm 3$ kpc} \\
ISM A$_{\rm V}$     & \multicolumn{2}{c}{$24 - 32$}   \\
\cutinhead{Dust Shell Parameters}
& \orich Shell & \crich Shell \\
$r_{\rm d,~min}$ (pc) & 0.12    & 0.0038 \\
$r_{\rm d,~max}$ (pc) & 5.8\phn & 0.12\phn \phn \\
$\theta_{\rm incl}$ & \multicolumn{2}{c}{$45\pm5^{\circ}$} \\
$\tdust$ at $r_{\rm d,~min}$ (K) & 112 & 512 \\
$\tau_{10.3\um,~{\rm eq}}$   & 0.54 & 0.036\\
$\tau_{10.3\um,~{\rm pole}}$ & 0.15 & 0.036 \\
$\rho_{\rm pole}/\rho_{\rm eq}$ at $r_{\rm d,~min}$ & 0.5 & 1 \\
$\mdot_{\rm dust}$ ($\msun$/yr) & 6.8 $\times 10^{-6}$ &  
5.7 $\times 10^{-9}$ \\
$v_{\rm exp}$\tablenotemark{a} & 50 km s$^{-1}$ & 50 km s$^{-1}$ \\
$\tau_{\rm dyn}$ (yr)\tablenotemark{b} & $1.1 \times 10^{5}$ & 
$2 \times 10^{3}$ \\
$a$ & 0.01 $\um$ & 0.002 $\um$ \\
Composition\tablenotemark{c} & 
\multicolumn{1}{l}{$96\%$ Amorphous Silicates} & 
\multicolumn{1}{l}{$100\%$ Amorphous Carbon} \\
& \multicolumn{1}{l}{$\phn 4\%$ Alminum Oxides} &\\
\enddata 
\tablenotetext{a}{Typical $v_{\rm exp}$ for LBVs (e.g., \citealt{nota95}).}
\tablenotetext{b}{$\tau_{\rm dyn} = \rdust / v_{\rm exp}$}
\tablenotetext{c}{Abundance $\%$ by mass; 
References: 
amorphous silicates; \citet{ld93},
aluminum oxides; \citet{begemann97,harman94},
amorphous carbon; \citet{zubko96}.}
\end{deluxetable}

\clearpage
\begin{deluxetable}{lcccc}
\tablecolumns{5} 
\tablewidth{0pc} 
\tablecaption{{\it JHK'} Photometry of \iras 18576+0341} 
\tablehead{ 
\colhead{Date} & 
\colhead{{\it J}}   & 
\colhead{{\it H}}    & 
\colhead{{\it K'}} &
\colhead{Ref.}} 
\startdata 
1989 May$-$Jun & 
12.0\phn$\pm$0.3\phn\phn & \phn8.99$\pm$0.03\phn & 6.91$\pm$0.02\phn & 
1 \\
1999 Jun 12 & 
12.21$\pm$0.12\phn & \phn8.94$\pm$0.06\phn & 7.07$\pm$0.05\phn & 
2 \\
1999 Nov 14  & 
12.44$\pm$0.07\phn & \phn9.32$\pm$0.05\phn & 7.60$\pm$0.08\phn & 
2 \\
2000 Aug ~5 & 
12.61$\pm$0.51\phn & 10.44$\pm$0.16\phn & 8.58$\pm$0.11\phn & 
2 \\
\enddata
\tablerefs{1. \citet{gmpp97}, 2. NIRIM observations.} 
\end{deluxetable}


\begin{thebibliography}{dummy} 
\bibitem[Allamandola, Hudgins, \& Sandford (1999)]{ahs99}
  Allamandola, L. J., Hudgins, D. M., \& Sandford, S. A.
  1999, \apjl, 511, L115

\bibitem[Becker et al.\/(1994)]{bwhz94}
  Becker, R. H., White, R. L., Helfand D. J., \& Zoonematkermani, S. 
  1994, \apjs, 91, 347

\bibitem[Begemann et al.\/(1997)]{begemann97}
  Begemann, B., Dorschner, J., Henning, TH., Mutschke, H., 
  G\"urtler. J., K\"ompe, C., \& Nass, R.
  1997, \apj, 476, 199
  
\bibitem[Beintema et al.\/(1996)]{beintema96}
  Beintema, D. A., van den Ancker, M. E., Molster, F. J., 
  Waters, L. B. F. M., Tielens, A. G. G. M., Waelkens, C., 
  de Jong, T., de Graauw, Th., Justtanont, K., Yamamura, I., 
  Heras, A., Lahuis, F., \& Salama, A.
  1996, \aap, 315, L369	  

\bibitem[Casali \& Hawarden (1992)]{casali92}
  Casali, M. M. \& Hawarden, T. G. 
  1992, JCMT UKIRT Newsletter, 4, 33

\bibitem[Chu, Weis, \& Garnett (1999)]{chu99}
  Chu, Y.-H., Weis, K., \& Garnett, D. R.
  1999, \aj, 117, 1433

\bibitem[Cohen \& Davies (1995)]{cohen95}
  Cohen, M. \& Davies J. K.
  1995, \mnras, 276, 715

\bibitem[Condon, Kaplan, \& Terzian (1999)]{ckt99}
  Condon, J. J., Kaplan, D. L., \& Terzian, Y.
  1999, \apjs, 123, 219

\bibitem[Damineli et al.\/(2000)]{damineli00}
  Damineli, A., Kaufer, A., Wolf, B., Stahl, O., Lopes, D. F.,
  \& de Ara{\' n}jo, F. X.
  2000, \apjl, 528, L101

\bibitem[Dayal et al.\/(1998)]{dayal98}
  Dayal, A., Hoffmann, W. F., Bieging, J. H., Hora, J. L., 
  Deutsch, L. K., \& Fazio, G. G. 
  1998, \apj, 492, 603

\bibitem[de Graauw et al.\/(1996)]{degraauw96}
  de Graauw, T. et al.
  1996, \aap, 315, L49

\bibitem[Egan et al. (1999)]{egan99}
  Egan, M. P., Price, S. D., Moshir, M. M., Cohen, M., 
  Tedesco, E. F., Murdock, T. L., Zweill, A., Burdick, S., 
  Bonito, N., Gugliotti, G. M., \& Duszlak, J. 
  1999, The Midcourse Space Experiment Point Source Catalog,
  Version 1.2 Explanatory Guide, 
  Air Force Research Laboratory Technical Report, AFRL-VS-TR-1999-1522

\bibitem[Elias et al.\/(1992)]{elias82}
  Elias, J. H., Frogel, J. A., Matthews, K., \& Neugebauer, G.
  1982, \aj, 87,m 1029

\bibitem[Garc\'{\i}a-Lario et al.\/(1997)]{gmpp97}
  Garc\'{\i}a-Lario, P., Manchado, A., Pych, W., Pottasch, S. R.
  1997, \aaps, 126, 479

\bibitem[Harman, Ninomiya, \& Adachi (1994)]{harman94}
  Harman, A., K., Ninomiya, S., \& Adachi, S.
  1994, J. Appl. Phys., 76, 8032

\bibitem[Hawkins et al.(1995)]{hawkins95}
  Hawkins, G. W., Skinner, C. J., Meixner, M., Jernigan, J. G., 
  Arens, J. F., Keto, E., \& Graham, J. R.
  1995, \apj, 452, 314

\bibitem[Hoffmann et al.\/(1998)]{hoffmann98}
  Hoffmann, W. F., Hora, J. L., Fazio, G. G., 
  Deutsch, L. K.  \&  Dayal, A. 
  1998,  in Infrared Astronomical Instrumentation, 
  ed. A. M. Fowler, Proc. SPIE 3354, 647, 658

\bibitem[Hrivnak, Volk, \& Kwok (2000)]{hvk00}
  Hrivnak, B. J., Volk, K., \& Kwok, S.
  2000, \apj, 535, 275

\bibitem[Humphreys \& Davidson (1979)]{humphreys79}
  Humphreys, R. M. \& Davidson, K.
  1994, \apj, 232, 409

\bibitem[Humphreys \& Davidson (1994)]{humphreys94}
  Humphreys, R. M. \& Davidson, K.
  1994, \pasp, 106, 1025

\bibitem[{\sl IRAS} Explanatory Supplement (1988)]{iras}
  Joint IRAS Science Working Group
  1988, Infrared Astronomical Satellite Catalogs and Atlases, 
  Version 2. Explanatory Suppliment,
  eds. C. A. Beichmen, G. Neugebauer, H. J. Habing, P. E. Clegg, \&
  T. J. Chester (Washington, DC: NASA Reference Publication)

\bibitem[Kessler, et al.\/(1996)]{kessler96}
  Kessler, M. F., Steinz, J. A., Anderegg, M. E., Clavel, J.,
  Drechsel, G., Estaria, P., Faelker, J., Riedinger, J. R.,
  Robson, A., Taylor, B. G., \& Ximenez de Ferran, S.
  1996, \aap, 315, L27

\bibitem[Kwok, Volk, \& Bidelman (1997)]{kvb97}
  Kwok, S., Volk, K., \& Bidelman, W.
  1997, \apjs, 112, 557

\bibitem[Lamers et al.\/(1996)]{lamers96}
  Lamers, H. J. G. L. M., Morris, P. W., Voors, R. H. M., 
  van Gent, J. I., Waters, L. B. F. M., de Graauw, Th.,
  Kudritzki, R. P., Najarro, F., Salama, A., \& Heras, A. M.
  1996, \aap, 315, L225

\bibitem[Langer, Garc{\' \i}a-Segura, \& Mac Low (1999)]{langer99}
  Langer, N., Garc{\' \i}a-Segura, G., \& Mac Low, M.-M.
  1999, \apjl, 520, L49

\bibitem[Laor \& Draine (1993)]{ld93}
  Laor, A. \& Draine, B. T.
  1993, \apj, 402, 441

\bibitem[Mastrodemos \& Morris (1999)]{mm99}
  Mastrodemos, N. \& Morris, M.
  1999, \apj, 523, 357

\bibitem[McLean \& Teplitz\/(1996)]{mclean96}
  McLean, I. S. \& Teplitz, H.
  1996, \aj, 112, 2500

\bibitem[Meixner et al.\/(1997)]{meixner97}
  Meixner, M., Skinner, C. J., Graham, J. R., Keto, E., 
  Jernigan, J. G., \& Arens, J. F.
  1997, \apj, 482, 897

\bibitem[Meixner et al.\/(1999)]{meixner99}
  Meixner, M., Ueta, T., Dayal, A., Hora, J. H., Fazio, G., 
  Hrivnak, B. J., Skinner, C. J., Hoffman, W. F., \&
  Deutsch, L. K.
  1999, \apjs, 122, 221

\bibitem[Meixner, Young Owl, \& Leach (1999)]{myol99}
  Meixner, M., Young Owl, L., \& Leach, R.
  1999, \pasp, 111, 997

\bibitem[Molster et al.\/(1996)]{molster96}
  Molster, F. J., van den Ancker, M. E., Tielens, A. G. G. M., 
  Waters, L. B. F. M., Beintema, D. A., Waelkens, C., 
  de Jong, T., de Graauw, Th., Justtanont, K., Yamamura, I., 
  Vandenbussche, B., \& Heras, A.
  1996, \aap, 315, L373

\bibitem[Morris et al.\/(1999)]{morris99}
  Morris, P. W., Waters, L. B. F. M., Barlow, M. J.,
  Lim, T., de Koter, A., Voors, R. H. M., Cox, P., 
  de Graauw, Th., Henning, Th., Hony, S., Lamers, H. J. G. L. M.,
  Mutschke, H., \& Trams, N. R.
  1999, Nature, 402, 502

\bibitem[Nota et al.\/(1995)]{nota95}
  Nota, A., Livio, M., Clampin, M., Schlte-Landbeck, R.
  1995, \apj, 448, 788

\bibitem[Nota et al.\/(1996)]{nota96}
  Nota, A., Pasquali, A., Clampin, M., Pollacco, D., \& Scuderi, S.
  1996, \apj, 473, 946

\bibitem[Omont (1986)]{omont86}
  Omont, A.
  1986, \aap, 164, 159

\bibitem[Parenago (1940)]{parenago40}
  Parenago, P. P.
  1940, Astron. Zhur., 17, 3

\bibitem[Polomski et al.\/(1999)]{polomski99}
  Polomski, E. F., Telesco, C. M., Pia, R K.,
  \& Fisher, R. S.
  1999, \aj, 118, 2369

\bibitem[Price \& Murdock (1983)]{pm83}
  Price, S. D. \& Murdock, T. L. 
  1983, The Revised AFGL Infrared Sky Survey Catalog, 
  AFGL-TR-83-0161 (Hanscom AFB, MA: Air Force Geophysics Laboratory,
  Air Force Systems Command, USAF)

\bibitem[Price, Murdock, \& Shivanandan (1981)]{pms81}
  Price, S. D., Murdock, T. L. \& Shivanandan, K.
  1981, in Infrared astronomy - Scientific/military thrusts and 
  instrumentation,
  ed. Howard J. Stears \& Nancy W. Boggess, SPIE, 280. 33

\bibitem[Robberto \& Herbst (1998)]{rh98}
  Robberto, M., \& Herbst, T. M.
  1998, \apj, 498, 400

\bibitem[Schlegel, Finkbeiner, \& Davis (1998)]{schlegel98}
  Schlegel, D. J., Finkbeiner, D. P. \& Davis, M.
  1998, \apj, 500, 525

\bibitem[Skinner et al.\/(1997)]{skinner97}
  Skinner, C. J., Meixner, M., Barlow, M. J., Collison, A. J.,
  Justtanont, K., Blanco, P., Pina, R., Ball, J. R., Keto, E.,
  Arens, J. F., \& Jernigan, J. G.
  1997, \aap, 3328, 290

\bibitem[Speck (1998)]{speck98}
  Speck, A. K.
  1998, PhD Thesis, University College London

\bibitem[Trams, Waters, \& Voors (1996)]{trams96}
  Trams, N. R., Waters, L. B. F. M., \& Voors, R. H. M.
  1996, \aap, 315, L213

\bibitem[Ueta, Meixner, \& Bobrowsky (2000)]{ueta00}
  Ueta, T., Meixner, M., \& Bobrowsky, M.
  2000, \apj, 528, 861
  
\bibitem[Van der Veen \& Habing (1988)]{vh88}
  Van der Veen, W. E. C. J. \& Habing, H. J.
  1988, \aap, 194, 125

\bibitem[Volk \& Cohen (1989)]{vc89}
  Volk, K. \& Cohen, M.
  1989, \aj, 98, 931

\bibitem[Voors et al.\/(1999)]{voors99}
  Voors, R. H. M., Waters, L. B. F. M., Morris, P. W.,
  Trams, N. R., de Koter, A. \& Bouwman, J.
  1999, 341, L67

\bibitem[Waters et al.\/(1997)]{waters97}
  Waters, L. B. F. M., Morris, P. W., Voors, R. H. M., 
  \& Lamers, H. J. G. L. M.
  1997, in Luminous Blue Variables: Massive Stars in Transition
  eds. A. Nota \& H. J. G. L. M. Lamers, ASP Conferenc Series,
  120, 326

\bibitem[White, Becker, \& Helfand (1991)]{wbh91}
  White, R. L., Becker, R. H., \& Helfand D. J.
  1991, \apj, 371, 148

\bibitem[Zijlstra et al.\/(1989)]{zijlstra89}
  Zijlstra, A. A., te Lintel Hekkert, P., Pottasch, S. R., 
  Caswell, J. L., Ratag, M., \& Habing, H. J.
  1989, \aap, 217, 157
   
\bibitem[Zubko et al.\/(1996)]{zubko96}
  Zubko, V. G., Mennella, V., Colangeli, L., \& Bussoletti, E.
  1993, \mnras, 282, 1321

\bibitem[Zuckerman \& Dyck (1986)]{zd86}
  Zuckerman, B. \& Dyck, H. M.
  1986, \apj, 311, 345

\bibitem[Zuckerman \& Lo (1987)]{zl87}
  Zuckerman, B. \& Lo, K. Y.
  1987, \aap, 173, 263

\end{thebibliography}
\end{document}